\documentclass[11pt]{pazh}

\usepackage[T2A]{fontenc}
\usepackage[utf8]{inputenc}

\usepackage{latexsym}
\usepackage{amssymb}

\usepackage{graphics}
\usepackage{graphicx}
\usepackage{latexsym}
\usepackage{amssymb}
\usepackage{amsmath}
\DeclareMathOperator{\sech}{sech}

\begin{document}

\setcounter{page}{1}

\sloppypar

\title{\bf Polar-bulge galaxies}

\author{V.P. Reshetnikov\inst{1}, S.S. Savchenko\inst{1}, A.V. Mosenkov\inst{1,2,3},
N.Ya. Sotnikova\inst{1}, D.V. Bizyaev\inst{4,5}} 

\institute{St.Petersburg State University, Universitetskii pr. 28, St.Petersburg, 
198504 Russia
\and
Sterrenkundig Observatorium, Universiteit Gent, Belgium
\and 
Central (Pulkovo) Astronomical Observatory of RAS, Russia
\and
Apache Point Observatory and New Mexico State University, USA
\and
Sternberg Astronomical Institute, Moscow State University, Russia
}

%\authorrunning{ }
\titlerunning{Polar bulges}

\abstract{Based on SDSS data, we have selected a sample of nine edge-on spiral galaxies 
with bulges whose major axes show a high inclination to the disk plane. Such objects 
are called polar-bulge galaxies. They are similar in their morphology to polar-ring 
galaxies, but the central objects in them have small size and low luminosity. We have 
performed a photometric analysis of the galaxies in the $g$ and $r$ bands and
determined the main characteristics of their bulges and disks. We show that the disks 
of such galaxies are typical for the disks of spiral galaxies of late morphological 
types. The integrated characteristics of their bulges are similar to the parameters 
of normal bulges. The stellar disks of polar-bulge galaxies often show large-scale warps, 
which can be explained by their interaction with neighboring galaxies or external
accretion from outside.
}
\titlerunning{Polar bulges}
\maketitle

\section{Introduction}

The bulge is one of the main subsystems of spiral
galaxies. Bulges distinguish from the other galactic subsystems
(disks, bars, spiral arms, etc.) by their brightness
distribution, shape, stellar population,
the pattern of stellar motion, and other characteristics.
For a long time the bulges were considered
as a kind of small elliptical galaxies surrounded
by stellar disks. In the last couple of decades this
simplified picture has changed to a more complex one:
it turned out that the bulges are not a homogeneous
type of objects. They break up into at least two different
groups: the classical bulges that resemble elliptical
galaxies in many respects (see, e.g., Renzini, 1999)
and the pseudo-bulges that are closer to galactic disks 
in a number of characteristics (Kormendy and Kennicutt
2004; Kormendy 2015). Athanassoula (2005)
introduced the third type,
boxy/peanut bulges, that are actually edge-on bars. The
situation is further complicated by the fact that different
types of bulges can coexist in the same galaxy (see,
e.g., Erwin et al. 2015).

In this paper we consider yet another type of
bulges, the bulges whose apparent major axis is highly
inclined to the galactic midplane. The
objects with such bulges are similar in their morphology
to the polar-ring galaxies whose polar component
dominates in luminosity. Corsini et al. (2012)
called such bulges {\it polar}. The polar bulges are extremely
rare. In a recent review devoted to the shape
of galactic bulges, M\'endez-Abreu (2015) mentions
only three such galaxies: NGC 4698 (Bertola et al.
1999; Corsini et al. 2012), NGC 4672 (Sarzi et al.
2000), and UGC 10043 (Matthews and de Grijs
2004). One of the reasons for the small number
of known polar bulges is that they can be identified
only in the case where the galactic disk is close to the
edge-on orientation to the line of sight and therefore,
the bulge inclination to the disk plane can be easily
detected. If this is not the case, then the bulge
looks like a bar and does not attract the attention of
researchers.

Given the small number of known objects,
the systematics of properties and the origin of the polar
bulges remain puzzles. In two of the three galaxies
(NGC 4698 and NGC 4672) the morphologically
decoupled bulges are known to be related to 
the kinematically decoupled subsystems in their cores
that  rotate almost orthogonally to the galactic disks
(Bertola et al. 1999; Sarzi et al. 2000). No kinematically
decoupled core was detected in UGC 10043
(Matthews and de Grijs 2004), although as the authors
note, this could be hindered by the limited spatial
resolution of the spectra they used and by the
powerful absorption stripe that shields the core.

The existence of kinematically and morphologically
decoupled structures in galaxies is usually associated
with a ``secondary'' event in their history. To
explain the formation of polar bulges, Matthews and
de Grijs (2004) considered three possible ``secondary''
events: the capture of a disk by a pre-existing ``naked''
bulge (this scenario resembles the formation of polar-ring
galaxies with the external accretion; Reshetnikov
and Sotnikova 1997); the capture of a 
small elliptical galaxy by a spiral galaxy (this mechanism
resembles the formation of collisional rings, but
the relative velocity of the galaxies must be small;
Appleton and Struck-Marcell 1996); and merging
of two disk galaxies (Bekki 1998). Qualitatively, all
these scenarios are capable of describing a number
of peculiarities of polar-bulge galaxies, and extensive
observational data are needed to choose between
them. First of all, it is necessary to 
increase the sample of known objects of this type
and to perform their detailed observational study and
simulations.

In this paper, we present a small sample
of new candidates to the polar-bulge galaxies and some
results of their photometric analysis. 
All numerical values in the paper
are given for the cosmological model with the Hubble
constant of 70 km s$^{-1}$ Mpc$^{-1}$ and 
$\Omega_m=0.3$, $\Omega_{\Lambda}=0.7$.

\section{The sample of polar-bulge galaxies and their analysis}

\subsection{The sample of galaxies}

In order to search for galaxies with morphologically decoupled
bulges we examined images of galaxies
from the EGIS (Edge-on Galaxies In SDSS) catalog
(Bizyaev et al. 2014), which includes almost
six thousand edge-on galaxies. Our examination
revealed more than twenty objects whose bulges visually
appeared elongated along the galaxy minor
axis. As the next step, we performed a photometric decomposition
of the images of all selected galaxies and 
estimated parameters of the
disks and bulges, including their apparent flattening
and position angles of the apparent major axes (see
the next section for the details of our analysis). 
Our final sample of nine candidates to the polar-bulge galaxies
includes only those objects whose bulges in the $g$
and $r$ bands showed an apparent flattening $b/a < 0.9$,
while the difference between the position angles of the
major axes for the disks and bulges exceeded 30$^{\rm o}$. 
The accuracy of the model parameters (see below) does
not allow us to reliably judge whether the bulges with
$b/a \geq 0.9$ can belong to the polar ones.

Table 1 is based of the SDSS\footnote{http://www.sdss.org} 
and NED\footnote{http://ned.ipac.caltech.edu} data, and summarizes
the main characteristics of the final sample
of the polar-bulge galaxies. Their $r$-band contour maps are shown in
Fig. 1. As it can be seen from the table, two galaxies
(\#1 and \#5) were previously included in the catalog
of candidates to the polar-ring galaxies (SPRC; Moiseev
et al. 2011) due to their unusual morphology. The
galaxy UGC 10043 (no. 7 in Table 1) was studied by
Matthews and de Grijs (2004) (see the Introduction).

\begin{table*}
\caption{Candidates for the polar-bulge galaxies}
\begin{center}
\begin{tabular}{|c|c|c|c|c|c|}
\hline
         &           &              &      &            &     \\
\#        & SDSS name & Alternative name & Type & $r$ (mag)  &  $z$  \\
         &           &              &      &            &     \\
\hline
         &           &              &      &            &     \\
         1  &  SDSS J015858.39-002923.2     &  SPRC-77 &     & 17.53 & 0.08099 \\
         2  &  ~~~~~~~~~~J095346.69+351456.8&  &     & 16.62 & 0.03928 \\
         3  &  ~~~~~~~~~~J102343.67+421917.9&  &     & 15.59 & 0.04611 \\
         4  &  ~~~~~~~~~J124331.21-142015.8 &  FGC~1494, RFGC~2365 &  Sc & 16.72 & \\
         5  &  ~~~~~~~~~~J133904.58+020949.5&  UGC~8634, FGC~1649, &  Sb & 15.10 & 0.02337 \\
            &                               &  RFGC~2619, SPRC-42 & & & \\
         6  &  ~~~~~~~~~~J151223.37+013823.9&  &     & 16.58 & 0.02850 \\
         7  &  ~~~~~~~~~~J154841.10+215208.7&  UGC~10043, FGC~1953, & Sbc & 14.34 & 0.00721 \\
            &                               &  RFGC~3043 & & & \\
         8  &  ~~~~~~~~~~J160028.24+143201.2&  &     & 15.96 & 0.03413 \\
         9  &  ~~~~~~~~~~J222006.54+253621.0&  &     & 16.96 & 0.04164 \\
            &        &              &       &           &  \\
\hline
\end{tabular}
\end{center}
\end{table*}

\begin{figure}
\centering
\includegraphics[width=7.4cm, angle=0, clip=]{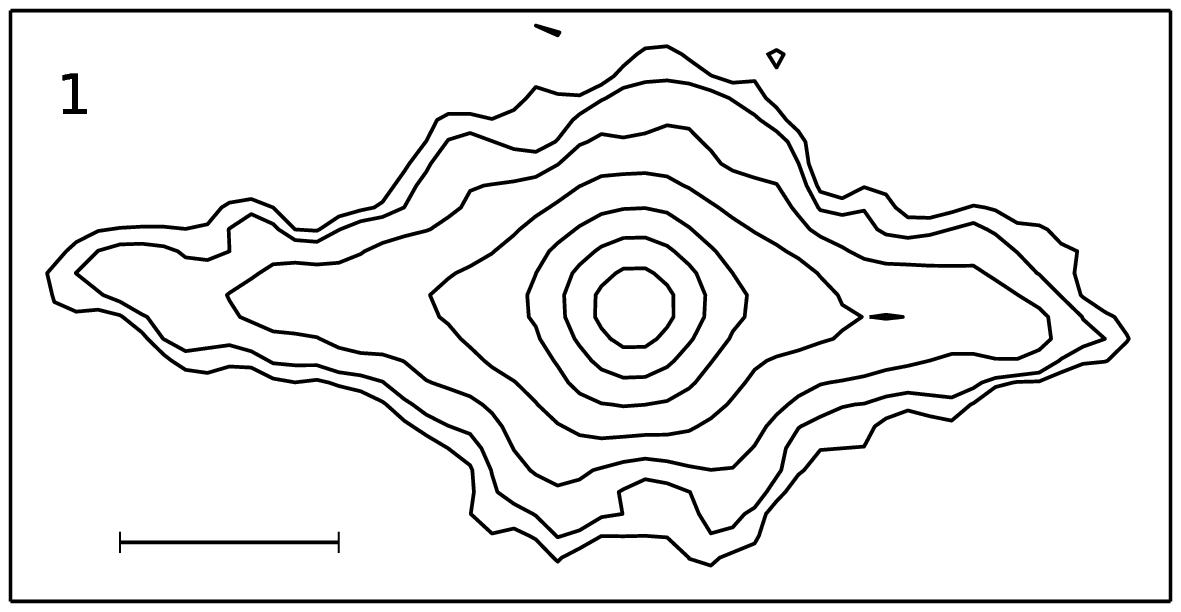}\\
\includegraphics[width=7.4cm, angle=0, clip=]{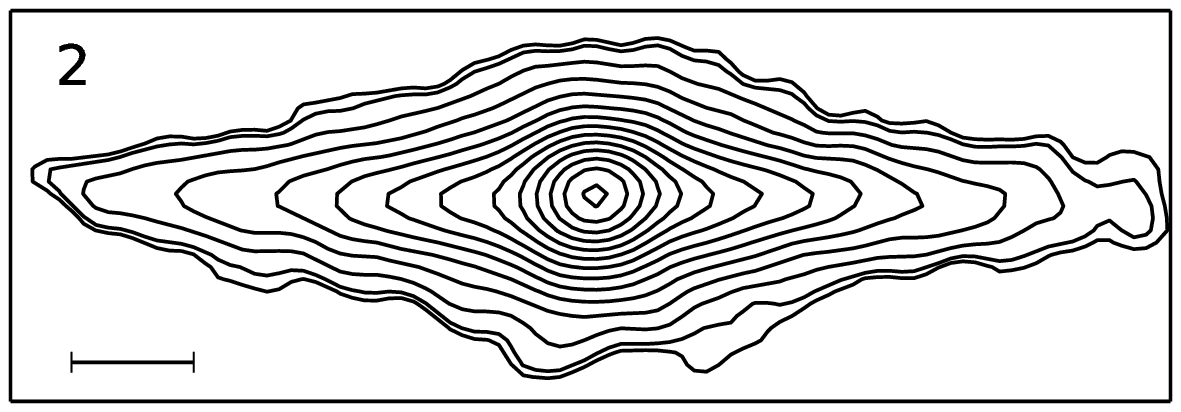}\\
\includegraphics[width=7.4cm, angle=0, clip=]{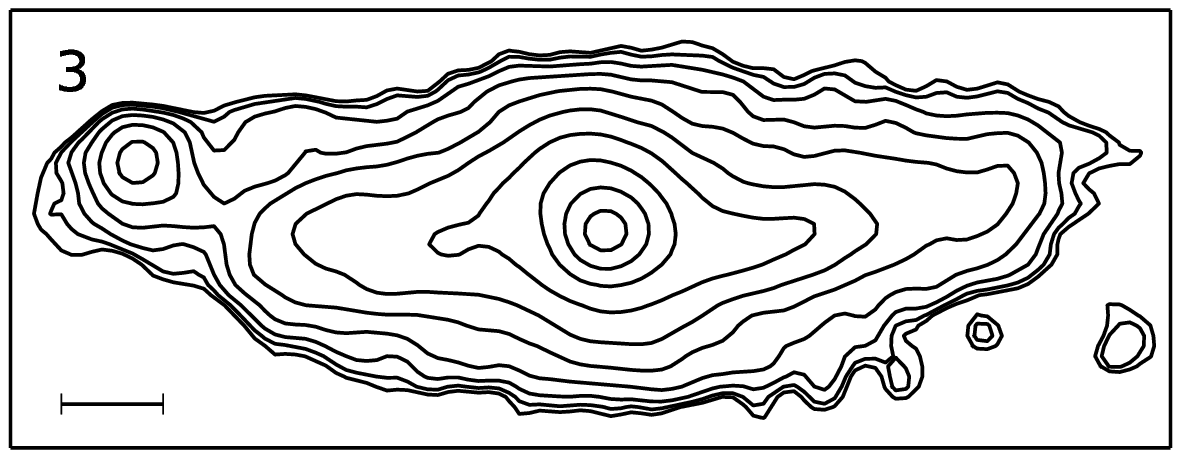}\\
\includegraphics[width=7.4cm, angle=0, clip=]{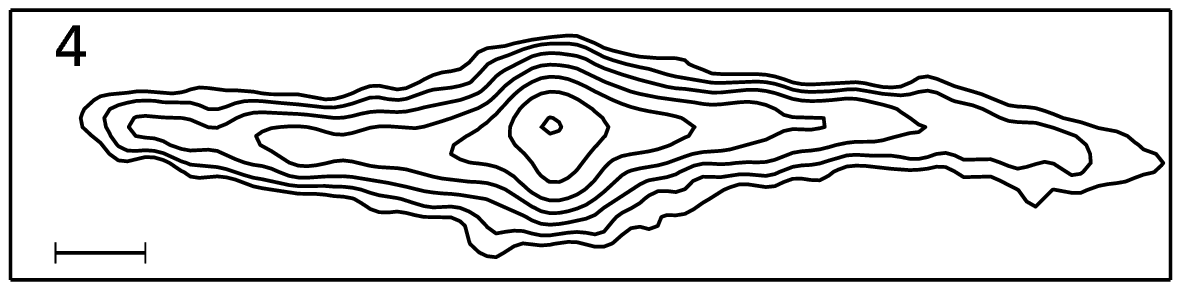}\\
\includegraphics[width=7.4cm, angle=0, clip=]{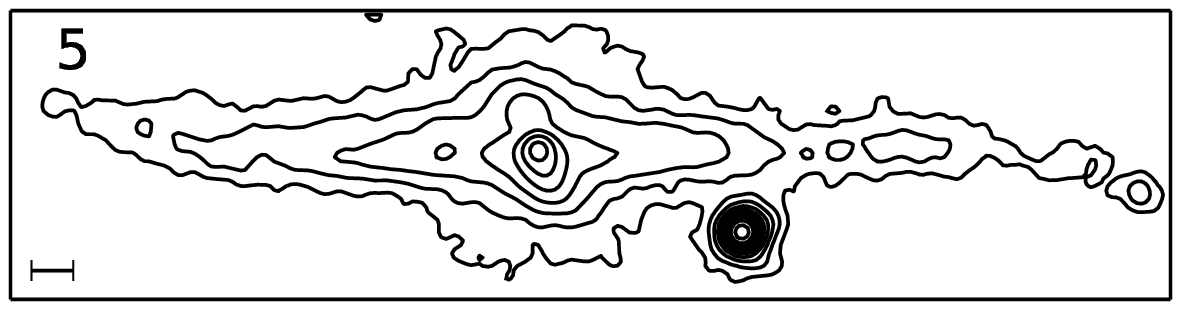}\\
\includegraphics[width=7.4cm, angle=0, clip=]{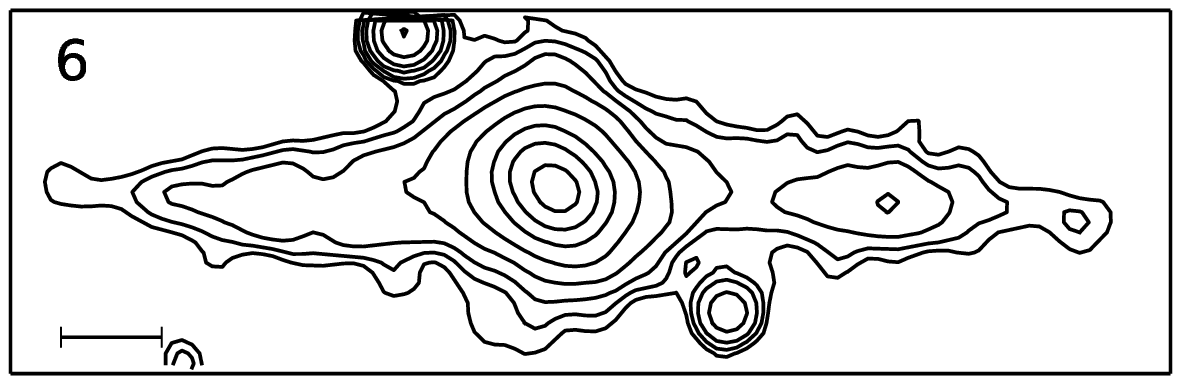}\\
\includegraphics[width=7.4cm, angle=0, clip=]{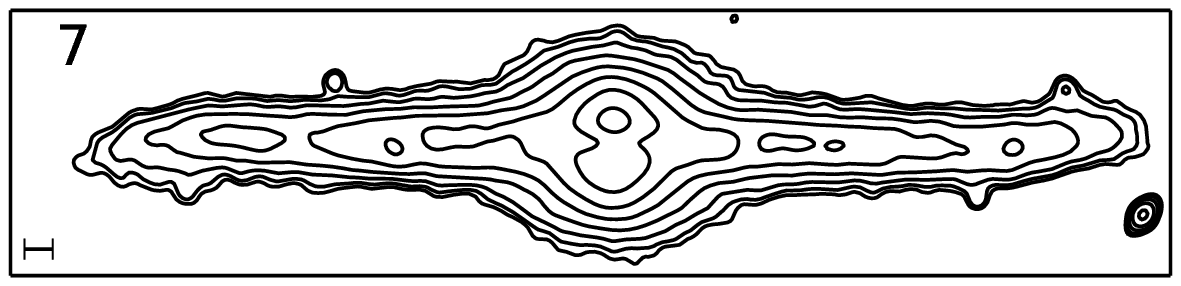}\\
\includegraphics[width=7.4cm, angle=0, clip=]{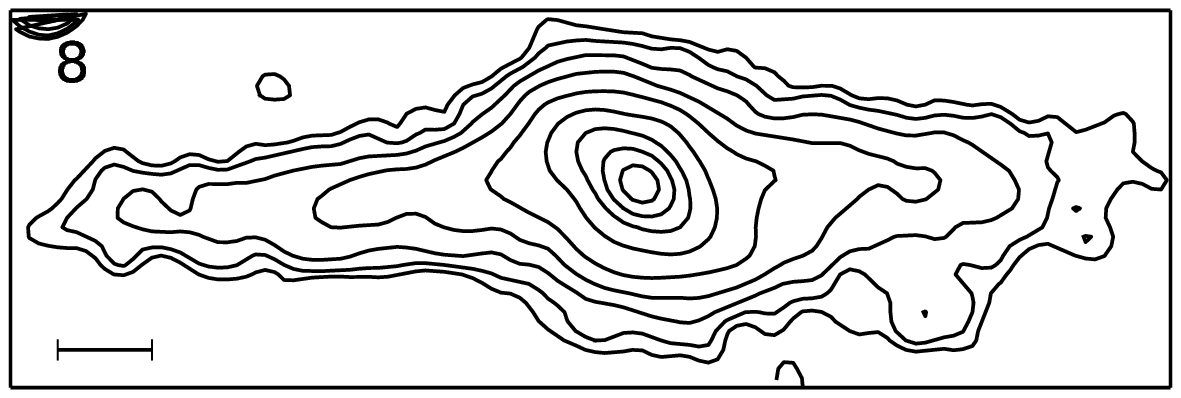}\\
\includegraphics[width=7.4cm, angle=0, clip=]{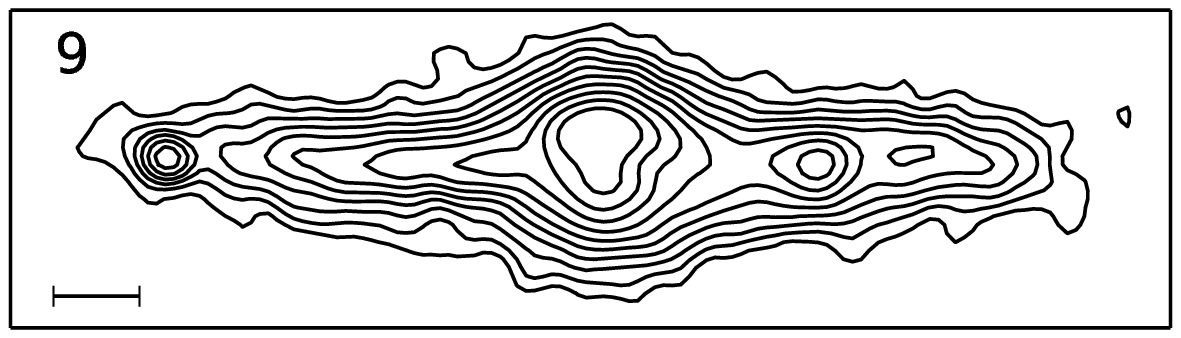}\\
\caption{The $r$-band contour maps of the galaxies with morphologically 
decoupled bulges. The contours were selected in a way to
highlight the global structure of each galaxy. The galaxy number from 
Table 1 is indicated in the upper left corner of each panel.
The length of the bar in the lower left corner is 10 arcsec. }
\end{figure}

\subsection{Photometric analysis}

To determine the photometric and structural parameters
of the considered galaxies, we decomposed
their $g$- and $r$-band images within the framework
of the two-component ``bulge + disk'' model.
The galaxy images were taken from the SDSS archive (SDSS
DR12; see Alam et al. 2015).

To describe the surface brightness distribution in
the bulge, we used the Sersic (1963) law:

%%%%%%%%%%%%%%%%%%%%%%%%%%%%%%%%%%%%%%%%%%%%%%%%%%%%%%%%%%%%%%%%%%%%
\begin{equation}
I(r) = 
I_{0,\mathrm{b}} \exp 
  \left[-\nu_n\left(\frac{r}{r_e}\right)^{\frac{1}{n}}\right],
  % \mu(r) = \mu_0 + \frac{2.5\nu_n}{\ln 10} \left( \frac{r}{r_e} \right)^{\frac{1}{n}},
\label{eqSersicMu}
\end{equation}
%%%%%%%%%%%%%%%%%%%%%%%%%%%%%%%%%%%%%%%%%%%%%%%%%%%%%%%%%%%%%%%%%%%%
where $I_{0,\mathrm{b}}$, $r_\mathrm{e}$ and $n$ are the central 
surface brightness, effective radius, and Sersic parameter, respectively;
$\nu_n$ is an $n$-dependent constant chosen in such
a way that a half of the total bulge luminosity was
encompassed within the $r_e$. The standard model of an edge-on
disk (van der Kruit and Searle 1981) was used as
the disk model:

%%%%%%%%%%%%%%%%%%%%%%%%%%%%%%%%%%%%%%%%%%%%%%%%%%%%%%%%%%%%%%%%%%%%
\begin{equation}
I(r, z) = 
I_{0,\mathrm{d}} 
  \left(\frac{r}{h}\right) K_1 
    \left(\frac{r}{h}\right) \sech^2\left(\frac{z}{z_0}\right),
\label{eqDiscEOn}
    \end{equation}
%%%%%%%%%%%%%%%%%%%%%%%%%%%%%%%%%%%%%%%%%%%%%%%%%%%%%%%%%%%%%%%%
where $I_{0,\mathrm{d}}$, $h$ and $z_0$ 
are the central surface brightness,
radial scale length and vertical scale height of the disk, and
$K_1$ is a modified first-order Bessel function.

The decomposition, i.e., the search for optimal parameters
that ensured the minimal residual between
the model and observations, was performed according
to the following scheme. At the preliminary stage,
the preparation of images was made, including the
image rotation in such a way that the galaxy major
axis was oriented along the x-axis of the image (the
method from Martin-Navarro et al. (2012) was used
to find the galaxy position angle), the cropping of
the images, and the masking of background objects and
dust lanes in the galaxies (if they were visible). At
the next stage we obtained the first approximation
of the galaxy model using the Differential Evolution
algorithm from the "Imfit" package (Erwin 2015). The
advantage of this algorithm is that it does not require
specific initial conditions for the optimization process
and only the range of possible parameters is needed
for its operation. The parameters obtained in this
way were then improved using the gradient descent
algorithm implemented in the "galfit" package (Peng
et al. 2010). The output parameters from the "galfit" were
assumed as the final galaxy model parameters.

Since the galaxies in our sample have rather
small angular sizes, atmospheric seeing
strongly affects the parameters being determined
(e.g., Trujillo et al. 2001). To compensate for this
effect, we convolved the photometric model with the
point spread function (PSF) simulating atmospheric
seeing and compared the convolved model with the
observed image. 
Here, we used the Moffat function
(Moffat 1969) as the PSF. We determined the parameters
of the Moffat function by fitting isolated field
stars. As a rule, there were 10 - 20 such PSF stars in
the field of each galaxy, which allowed us to reliably
determine the PSF parameters and to estimate the
dispersion of these parameters.

To estimate the errors of the decomposition results,
we performed Monte Carlo simulations by assuming
that the main sources of errors were the image
noise and the discrepancy between the model and
real PSFs. For each galaxy, we constructed a model
corresponding to the parameters determined during
the decomposition, convolved it with the observed
PSF, and added noise whose value corresponded to
the real parameters. 
The PSF parameters differed from their
true values according to a normal law with a dispersion
equal to the dispersion of the parameters of individual
PSF stars obtained at the PSF construction
stage. For each galaxy, we repeated the decomposition procedure
250 times and obtained the mean values and the 
uncertainties of the decomposition
parameters.

The mean difference between the total magnitudes
(defined as the sum of the model bulge and disk) 
and the {\em modelMag} magnitudes from the SDSS is
$-0\fm37 \pm 0\fm29$  in the $g$ band and 
$-0\fm23 \pm 0\fm25$ in
the $r$ band. If we take into account the fact that
the SDSS photometric data are marked as unreliable
for four galaxies, then the agreement becomes better
for the remaining five galaxies: 
$-0\fm21 \pm 0\fm10$ and $-0\fm10 \pm 0\fm14$
in the same bands. In the subsequent
discussion, we use the results of our photometry.

The results of our photometric decomposition of the galaxies
from the sample in the $r$ band are shown in Table~2.
For each object, the first row of the table gives the
measured parameter and the second row gives the
measurement undertainty of this parameter estimated as
described above. The first column in Table 2 contains
the galaxy number according to Table 1. The
next four columns present the bulge parameters: the
effective surface brightness, effective radius, Sersic
index and apparent flattening. The next columns show the following characteristics of
the galactic disks: the observed central
surface brightness, exponential scale length, and vertical
scale height of the brightness distribution. The
observed difference between the position angles of the
major axes for the bulge and disk is given in the ninth
column of Table 2. (Note that $\Delta$P.A. in the table is
only the model-averaged difference of the position angles.
In practice, the position angles of the major axes
for the disks and bulges exhibit noticeable variations
for a number of galaxies.) The last three columns
give the galaxy absolute magnitude, its color index,
and the bulge-to-total 
luminosity ratio. The surface brightnesses, absolute magnitudes,
and color indices in Table 2 were corrected
for the foreground extinction in the Milky Way according to Schlafly and
Finkbeiner (2011).

\begin{table*}
\caption{Photometric characteristics of the galaxies in the $r$ band}
\begin{center}
\begin{tabular}{|c|c|c|c|c|c|c|c|c|c|c|c|}
\hline
N & $\mu_e$ &  $r_e$ & $n$ & $q$  & $\mu_0$ & $h$   & $z_0$ & $\Delta$P.A. & $M_r$ & $g - r$ & $B/T$ \\
  &         & ($''$) &     &      &         & ($''$)& ($''$)& ($^{\rm o}$) &       &       \\
  \hline               
  1 & 22.57   & 2.17    & 6.84& 0.85 &  22.16  & 4.25  & 0.88  &    73.0      & -20.51 & 0.75 &0.80  \\
    &  0.33   & 0.32    & 2.40& 0.04 &   0.10  & 0.17  & 0.06  &     3.6      &       &   &    \\
\hline
2 & 22.03   & 2.73    & 1.82& 0.87 &  20.93  & 5.09  & 1.21  &    74.7      & -19.85& 1.02 &0.36 \\
      &  0.44   & 0.49    & 2.23& 0.10 &   0.23  & 0.32  & 0.15  &    10.3      &       &  &    \\
  \hline
  3 & 21.46   & 2.17    & 5.44& 0.69 &  21.22  & 6.37  & 2.77  &    71.3      & -20.86& 0.56  &0.36 \\
    &  0.17   & 0.15    & 0.86& 0.03 &   0.02  & 0.07  & 0.03  &     1.8      &       &  &    \\
\hline
4 & 21.94   & 2.55    & 0.39& 0.47 &  21.31  & 6.48  & 1.60  &    89.5      &       & 0.58 &0.18 \\
  &  0.03   & 0.03    & 0.02& 0.01 &   0.01  & 0.06  & 0.02  &     0.6      &       &  &    \\
  \hline
  5 & 21.38   & 4.12    & 2.53& 0.42 &  21.39  & 12.00 & 3.10  &    72.2      & -20.07& 0.57 &0.36 \\
& 0.04    & 0.06    & 0.16& 0.02 &   0.01  &  0.09 & 0.02  &     0.4      &       &  &    \\
\hline
6 & 20.19   & 1.41    & 1.50& 0.66 &  22.47  & 8.22  & 2.43  &    49.8      & -19.09& 0.44 &0.56 \\
  & 0.13    & 0.07    & 0.08& 0.02 &   0.06  & 0.37  & 0.29  &     3.2      &       &  &    \\
  \hline
  7 & 20.89   & 5.95    & 0.86& 0.80 &  21.61  & 23.67 & 5.43  &    89.7      & -18.75& 0.64 &0.44 \\
    &  0.01   & 0.04    & 0.02& 0.01 &   0.01  &  0.08 & 0.02  &     0.1      &       &  &    \\
\hline
8 & 21.40   & 3.04    & 4.13& 0.42 &  21.88  & 8.28  & 2.97  &    39.1      & -20.09& 0.65 &0.47 \\
  &  0.06   & 0.08    & 0.48& 0.03 &   0.03  & 0.16  & 0.02  &     0.7      &       &  &    \\
  \hline
 9 & 22.07   & 2.69    & 0.72& 0.68 &  21.37  & 7.54  & 2.25  &    87.4      & -20.12& 0.57 &0.20 \\
&  0.36   & 0.23    & 0.08& 0.03 &   0.08  & 0.21  & 0.12  &     6.7      &       & &     \\
\hline
\end{tabular}
\end{center}
\end{table*}

\section{Results and discussion}

\subsection{General characteristics of the galaxies}

The integrated characteristics of the polar-bulge galaxies
in the $r$ band are as follows: the mean
absolute magnitude $< M > = -19.9 \pm 0.7$, the mean
color index $< g - r > = 0.64 \pm 0.16$, and the exponential
scale length of the disk $< h > = 5.1 \pm 1.0$ kpc. In
all these characteristics, our objects are typical bright
spiral galaxies of late morphological types (see, e.g.,
Bizyaev et al. 2014). The mean vertical scale height of
the stellar disk in the $r$ band, $< z_0 > = 1.50 \pm 0.54$ kpc,
is also near the peak of the distribution for the galaxies
from the EGIS catalog. If we exclude the two galaxies
(\#3 and \#6 in Table 1) whose inclination visually
deviates most strongly from 90$^\mathrm{o}$, then this value decreases
to $< z_0 > = 1.36 \pm 0.45$ kpc.

The mean relative disk thickness for the galaxies of
our sample is $< z_0 / h > = 0.285 \pm 0.07$; without galaxies
\#3 and \#6, it is $< z_0 / h > = 0.265 \pm 0.06$. These values
are close to the galactic disk thickness in well-defined
samples of edge-on spiral galaxies (Mosenkov et al. 2015).

The mean ratio of the disk exponential scale length in
the $g$ and $r$ bands for the considered sample is 1.09$\pm$0.10, 
which implyes the existence
of a color gradient: the galactic disks become bluer
towards the periphery. This feature is typical for
most of spiral galaxies.

An interesting structural peculiarity of the galaxies
in our sample is that their disks exhibit prominent
integral-shaped warps (see Fig. 1) in approximately
half of the cases. Since the warp is easier
to detect in the case where its line of nodes is close
to the line of sight, it can be concluded that the
stellar disks in most of the polar-bulge galaxies are warped.
Strong optical warps are observed in galaxies relatively
rarely. As a rule, their existence is associated
with the gravitational interaction of galaxies and
external accretion (Reshetnikov and Combes 1998,
1999; Ann and Park 2006).

We did not specially study the spatial environment
of the galaxies from our sample, but a simple analysis
of the SDSS images (in particular, deeper images
from Stripe 82 are available for galaxy 1 from our
list; Abazajian et al. 2009) showed that relatively
close galaxies of comparable luminosity and a number
of fainter companions are observed near most of the
objects (six out of nine). Unfortunately, there are
no available redshifts for the neighboring galaxies, 
which does not allow us to decide about their spatial
association with our objects.

More reliable data are available for previously
known galaxies with decoupled bulges. UGC 10043
(\#7 in Table 1) is a member of a group, and it
interacts with the galaxy MCG+04-37-035 
(Aguirre et al. 2009). Two other galaxies
(NGC 4672 and NGC 4698; see Introduction)
are not isolated either. NGC 4672 is a member
of a group of galaxies (Garcia 1993); NGC 4698 is a
member of the Virgo cluster of galaxies.

\subsection{Photometric characteristics of bulges}

The distributions of the main characteristics of
the decoupled bulges in the $r$ band are shown in Fig. 2.
By comparing these distributions with the characteristics
of the central objects in polar-ring galaxies (Fig. 2
in Reshetnikov and Combes 2015) it can be noted
that, as expected, the bulges are more
compact and faint, on average. For example, the mean absolute
$r$-band magnitude for the bulges of our sample is
$< M > = -18.95 \pm 0.77$ (compare with -20.34 for the centers
of the polar-ring galaxies), while the mean effective radius
in the same band is $< r_e > = 1.86 \pm 0.74$ kpc (3.2 kpc
for polar-ring galaxies). The distribution of the Sersic
index for the bulges appears almost flat (Fig. 2b) and
does not show a distinct peak typical for the bulges
of spiral galaxies of early morphological types (Tasca
and White 2011). In general, the observed distribution
is close to that for the bulges of the late-type 
spiral galaxies, with caution from the limited size of our sample. 

\begin{figure}
\centering
\includegraphics[width=16cm, angle=-90, clip=]{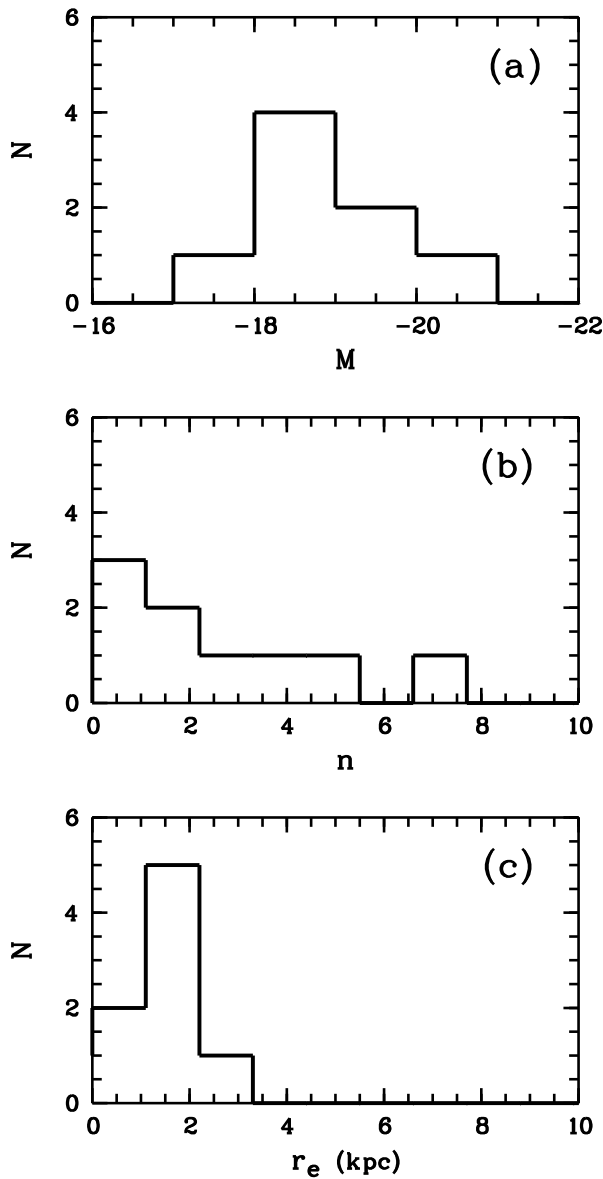}\\
\caption{Observed distributions of the polar-bulge characteristics in 
the $r$ band: the absolute magnitudes (a), Sersic index (b), and
effective radius (c).}
\end{figure}

The mean color index of the decoupled bulges,
$< g - r > = 0.88 \pm 0.13$, is considerably redder than
that of the surrounding disks, for which $< g - r > =
0.49 \pm 0.22$. Given that the colors of the edge-on
disks were not corrected for the internal extinction, the
color difference between the bulges and disks must
be even greater. This large color difference between
the bulges and disks is typical for galaxies of late
morphological types (Mollenhoff 2004).

Fig. 3 shows the photometric scaling relations
for the bulges. The relation between the luminosity
of morphologically decoupled bulges and their Sersic
index is displayed in Fig. 3a. The dashed line
in Fig. 3 corresponds to mean relation 
for the bulges of
almost 1000 galaxies constructed from the data of
Gadotti (2009) (see Eq. (25) and Fig. 17 in Mosenkov
et al. (2014)). 
As it can be seen from Fig. 3a, the bulges
we consider roughly follow the trend typical for the regular
galaxies.

\begin{figure}
\centering
\includegraphics[width=16.5cm, angle=-90, clip=]{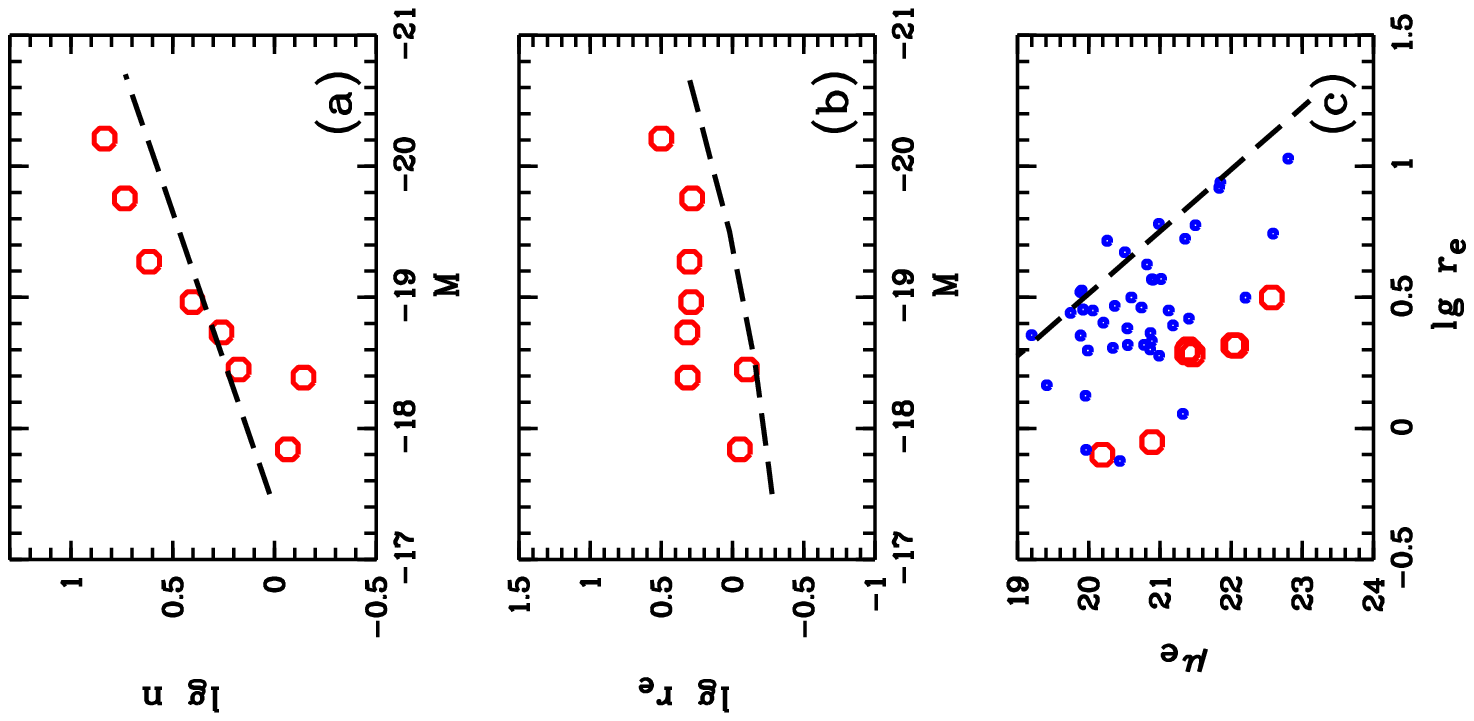}\\
\caption{The scaling relations for the polar bulges in the $r$ band (circles). 
(a) The luminosity vs Sersic index relation; the dashed
line indicates the mean relation for the bulges of regular galaxies (see text). 
(b) The luminosity vs effective radius (in kpc) relation;
the mean relation for normal bulges is indicated by the dashed line. 
(c) The Kormendy relation; the dashed line indicates the
relation for the E/S0 galaxies in the Coma cluster (Houghton et al. 2012); 
the dots are the characteristics of the polar-ring central 
galaxies (Reshetnikov and Combes 2015).}
\end{figure}

Fig. 3b illustrates the luminosity -- size relation
for the decoupled bulges in the comparison with the mean
relation for the bulges plotted from the data of
Gadotti (2009) (Eq. (31) in Mosenkov et al. (2014)).
The polar bulges follow the relation drawn for the regular bulges,
with a slight shift towards the larger sizes. It is difficult
to judge the significance of this shift because our
sample is small and in addition, the normal bulges
show a large scatter on the luminosity -- size diagram.

The position of the bulges in the Kormendy relation
are shown in Fig. 3c. The bulges are located
on this plane approximately along the canonical relation
for the E/S0 galaxies with a shift from it and from
the central polar-ring galaxies towards the smaller size
and down to the lower central brightness. 
There are faint early-type
galaxies and bulges (Capaccioli et al. 1992) in
this region (approximately at log\,$r_\mathrm{e} \leq 0.5$). The polar
bulges are at the edge of the region where the polar-ring
galaxies are located and are comparable in their
characteristics to the faintest and smallest host
objects in the polar rings (Fig. 3c).

\section{Conclusions}

We presented a sample of nine edge-on spiral
galaxies whose bulges were decoupled from their
disks in the sense that the major axes of the bulges are highly inclined
to the disk midplane. Note that we selected the objects
based on a purely morphological analysis; therefore,
there can be chance projections of two galaxies
among our objects. The galaxies with such decoupled,
or polar, bulges are extremely rare. Based on
the number of galaxies we examined when searching
for the polar bulges, we can roughly estimate their
relative fraction among the spiral galaxies as $\sim 10^{-3}$.

Our analysis of the photometric characteristics for
the galactic disks in the polar-bulge galaxies showed them to be similar to the
disks of ordinary late-type (Sc--Sd) spiral galaxies.
If we disregard the elongation along
the minor axes of the galaxies, the bulges also
appear quite typical. The polar
bulges are similar to the faintest and smallest of 
the central galaxies in the polar-ring galaxies.

The peculiarity of the polar-bulge galaxies is that
galaxies with strong stellar disk warps are encountered
relatively frequently among them. Since the warp
is often interpreted as a consequence of an external perturbation
and external accretion, these factors may
also be responsible for the formation of polar bulges.
As has been noted in Introduction, kinematically
decoupled subsystems rotating almost orthogonally
to the galactic disks are observed in two out of three
previously studied polar bulges. The existence of such
subsystems is usually associated with the external
accretion. Consequently, the detection of
such kinematically decoupled structures in the polar
bulges of other galaxies would give a good argument
for the hypothesis of the formation of decoupled
bulges under external accretion.

However, the question of what is the secondary
structure in the polar-bulge galaxies, their disks or bulges,
remains unclear. Their similarity to the polar-ring galaxies,
where the formation of polar disks/rings is well
reproduced in the models with the external accretion
from another galaxy or from a cloud of intergalactic gas
(Reshetnikov and Sotnikova 1997; Bournaud and
Combes 2003; Maccio et al. 2006), may suggest that
it is the disk in such objects that is secondary. From
this viewpoint, the polar-bulge galaxies are simply
the polar-ring galaxies with faint central galaxies. On
the other hand, as was discussed by Matthews and
de Grijs (2004), the opposite scenario, whereby a
pre-existing spiral galaxy captures a small elliptical
galaxy in the polar plane to become its bulge, is also
possible. Less exotic scenarios can also be
considered. In general, it is clear that new observing
data and realistic numerical simulations are necessary to
clarify the nature of these unique objects.

\bigskip
\section*{Acknowledgments}
This study was financially supported by the Russian
Foundation for Basic Research (project nos. 14-02-810 
and 13-02-00416). DB is supported by grant 
RSCF-14-22-00041.

\section*{REFERENCES}

\hspace{0.4cm}1.\,K.N.\,Abazajian, J.K.\,Adelman-McCarthy, M.A.\,Agueros, et al., Astrophys. J. Suppl. Ser.
182, 543 (2015).

2.\,P.\,Aguirre, J.M.\,Uson, and L.D.\,Matthews, Rev. Mex. Astron. Astrofis. 35, 201 (2009).

3.\,S.\,Alam, F.D.\,Albareti, C.A.\,Prieto, F.\,Anders,
F.S.\,Anderson, T.\,Anderton, B.H.\,Andrews, E.\,Armengaut,
et al., Astrophys. J. Suppl. Ser. 219, 12A (2015).

4.\,H.B.\,Ann and J.-C.\,Park, New Astron. 11, 293 (2006).

5.\,P.N.\,Appleton and C.\,Struck-Marcell, Fundam. Cosm. Phys. 16, 111 (1996).

6.\,E.\,Athanassoula, Mon. Not. R. Astron. Soc. 358, 1477 (2005).

7.\,K.\,Bekki, Astrophys. J. 499, 635 (1998).

8.\,F.\,Bertola, E.M.\,Corsini, J.C.\,Vega\,\,Beltran,
A.\,Pizzella, M.\,Sarzi, M.\,Cappellari, and S.J.\,Funes,
Astrophys. J. 519, L127 (1999).

9.\,D.V.\,Bizyaev, S.J.\,Kautsch, A.V.\,Mosenkov,
V.P.\,Reshetnikov, N.Ya.\,Sotnikova, N.V.\,Yablokova,
and R.W.\,Hillyer, Astrophys. J. 787, ID 24 (2014).

10.\,F.\,Bournaud and F.\,Combes, Astron. Astrophys. 401, 817 (2003).

11.\,M.\,Capaccioli, N.\,Caon, and M.\,D’Onofrio, Mon. Not. R. Astron. Soc. 259, 323 (1992).

12.\,E.M.\,Corsini, J.\,M\'endez-Abreu, N.\,Pastorello,
E.\,dalla Bonta, L.\,Morelli, A.\,Beifiori, A.\,Pizzella,
and F.\,Bertola, Mon. Not. R. Astron. Soc. 423, L79 (2012).

13.\,P.\,Erwin, Astrophys. J. 799, id. 226 (2015).

14.\,P.\,Erwin, R.P.\,Saglia, M.\,Fabricius, J.\,Thomas,
N.\,Nowak, S.\,Rusli, R.\,Bender, J.C.\,Vega Beltran,
and J.E.\,Beckman,Mon. Not. R. Astron. Soc. 446, 4039 (2015).

15.\,D.A.\,Gadotti, Mon. Not. R. Astron. Soc. 393, 1531 (2009).

16.\,A.M.\,Garcia, Astron. Astrophys. Suppl. Ser. 100, 47 (1993).

17.\,R.C.W.\,Houghton, R.L.\,Davies, E.\,dalla Bonita,
and R.\,Masters, Mon. Not. R. Astron. Soc. 423, 256 (2012).

18.\,J.\,Kormendy, Galactic Bulges, Ed. by E.\,Laurikainen,
R.F.\,Peletier, and D.A.\,Gadotti (Springer, New York, 2015).

19.\,J.\,Kormendy and R.C.\,Kennicutt, Ann. Rev. Astron. Astrophys. 42, 603 (2004).

20.\,P.C.\,van der Kruit and L.\,Searl, Astron. Astrophys. 95, 105 (1981).

21.\,A.V.\,Maccio, B.\,Moore, and J.\,Stadel, Astrophys. J. 636, L25 (2006).

22.\,I.\,Martin-Navarro, J.\,Bakos, I.\,Trujillo, J.H.\,Knapen,
E.\,Athanasoula, A.\,Bosma, S.\,Comeron, B.G.\,Elmegreen, et al., Mon. Not. R. Astron.
Soc. 427, 1102 (2012).

23.\,L.D.\,Matthews and R.\,de Grijs, Astron. J. 128, 137 (2004).

24.\,J.\,M\'endez-Abreu, Galactic Bulges, Ed. by E.\,Laurikainen,
R.F.\,Peletier, and D.A.\,Gadotti (Springer, New York, 2015).

25.\,A.F.J.\,Moffat, Astron. Astrophys. 3, 455 (1969).

26.\,A.V.\,Moiseev, K.I.\,Smirnova, A.A.\,Smirnova, and V.P.\,Reshetnikov, 
Mon.Not. R. Astron. Soc. 418, 244 (2011).

27.\,C.\,Mollenhoff, Astron. Astrophys. 415, 63 (2004).

28.\,A.V.\,Mosenkov, N.Ya.\,Sotnikova, and V.P.\,Reshetnikov,
Mon. Not. R. Astron. Soc. 441, 1066 (2014).

29.\,A.V.\,Mosenkov, N.Ya.\,Sotnikova, V.P.\,Reshet- nikov,
D.V.\,Bizyaev, and S.J.\,Kautsch, Mon. Not. R. Astron. Soc. 451, 2376 (2015).

30.\,Ch.Y.\,Peng, L.C.\,Ho, Ch.D.\,Impey, and H.-W.\,Rix,
Astron. J. 139, 2097 (2010).

31.\,A.\,Renzini, The Formation of Galactic Bulges, Ed. by C.M.\,Carollo, 
H.C.\,Ferguson, and R.F.G.\,Wyse (Cambridge Univ. Press, Cambridge, 1999).

32.\,V.\,Reshetnikov and F.\,Combes, Astron. Astrophys. 337, 9 (1998).

33.\,V.\,Reshetnikov and F.\,Combes, Astron. Astrophys. Suppl. Ser. 138, 101 (1999).

34.\,V.\,Reshetnikov and F.\,Combes, Mon. Not. R. Astron. Soc. 447, 2287 (2015).

35.\,V.\,Reshetnikov and N.\,Sotnikova, Astron. Astrophys. 325, 933 (1997).

36.\,M.\,Sarzi, E.M.\,Corsini, A.\,Pizzella, J.C.\,Vega Beltran, M.\,Cappellari, 
J.G.\,Funes, and F.\,Bertola, Astron. Astrophys. 360, 439 (2000).

37.\,E.F.\,Schlafly and D.P.\,Finkbeiner, Astrophys. J. 737, ID 103 (2011).

38.\,J.L.\,Sersic, Bol. Asoc. Argentina Astron. 6, 41 (1963).

39.\,L.A.M.\,Tasca and S.D.M.\,White, Astron. Astrophys. 530, A106 (2011).

40.\,I.\,Trujillo, J.A.L.\,Aguerri, J.\,Cepa, and C.M.\,Gutierrez, 
Mon. Not. R. Astron. Soc. 328, 977 (2001).

\end{document}